\documentclass[reprint,superscriptaddress,amsmath,amssymb,aps,pra]{revtex4-2}
\pdfoutput=1
\usepackage{graphicx}
\usepackage{dcolumn}
\usepackage{bm}
\usepackage{upgreek}

\begin{document}

\title{From Raman frequency combs to supercontinuum generation in nitrogen-filled hollow-core anti-resonant fiber}

\author{Shou-Fei Gao}
\affiliation{%
Guangdong Provincial Key Laboratory of Optical Fiber Sensing and Communication, Institute of Photonics Technology, Jinan University, Guangzhou, 511443, China
}%
\affiliation{%
 National Center of Laser Technology, Institute of Laser Engineering, Beijing University of Technology, Beijing 100124, China
}%
\author{Ying-Ying Wang}
\email{dearyingyingwang@hotmail.com}
\affiliation{%
Guangdong Provincial Key Laboratory of Optical Fiber Sensing and Communication, Institute of Photonics Technology, Jinan University, Guangzhou, 511443, China
}%
\affiliation{%
 National Center of Laser Technology, Institute of Laser Engineering, Beijing University of Technology, Beijing 100124, China
}%
\author{Federico Belli}
\author{Christian Brahms}
\affiliation{
 School of Engineering and Physical Sciences, Heriot-Watt University, Edinburgh EH14 4AS, United Kingdom
}%
\author{Pu Wang}
\affiliation{%
 National Center of Laser Technology, Institute of Laser Engineering, Beijing University of Technology, Beijing 100124, China
}%
\author{John C. Travers}
\email{j.travers@hw.ac.uk}
\affiliation{
 School of Engineering and Physical Sciences, Heriot-Watt University, Edinburgh EH14 4AS, United Kingdom
}%

\date{\today}

\begin{abstract}
 We demonstrate a route to supercontinuum generation in gas-filled hollow-core anti-resonant fibers through the creation of a broad vibrational Raman frequency comb followed by continuous broadening and merging of the comb lines through either rotational Raman scattering or the optical Kerr effect. Our demonstration experiments, utilizing a single pump pulse with 20~ps duration at 532~nm in a nitrogen-filled fiber, produce a supercontinuum spanning from 440~nm to 1200~nm, with an additional deep ultraviolet continuum from 250~nm to 360~nm. Numerical results suggest that this approach can produce even broader supercontinuum spectra extending from the ultraviolet to mid-infrared.
\end{abstract}

\maketitle

\section{Introduction}
Stimulated Raman scattering (SRS)---a process of cascaded inelastic scattering of laser light from molecules---can create broad sets of discrete spectral lines from narrowband driving lasers~\cite{imasaka1989,Harris1,baker_femtosecond_2011}. Such Raman frequency combs have been widely studied in both free-space \cite{Sokolov1,kawano_generation_1998} and waveguide geometries \cite{Benabid399,Couny1118} and found application in the synthesis of trains of light-field transients \cite{Kawano1997,chan_synthesis_2011,Katsuragawa2} and single ultrashort pulses \cite{Harris2,Korn} as well as for efficient wavelength conversion \cite{Benabid2004,zheng_freely_2015}. By tightly confining both the gas molecules and laser light inside hollow-core photonic-crystal fibers (HC-PCF), such as anti-resonant guiding fibers, Raman frequency combs can be efficiently generated directly from quantum noise fluctuations and do not require multi-color pumping~\cite{Benabid399,Couny1118,Wang2010,Abdolvand:12,Tani:15,Benoit:15}. However, the Raman combs produced to date, while spanning multiple octaves, have always consisted of discrete spectral lines with at most moderate spectral broadening or shifting~\cite{Tani:15, Strickland}.

In this paper we demonstrate, both experimentally and numerically, that SRS can create a broadband and smooth supercontinuum from a single narrowband pump laser. The key mechanism is a two-step process. At the start of the fiber, an extremely broad frequency comb is generated through a large vibrational frequency shift. At larger propagation distances this comb is subsequently broadened into a smooth supercontinuum by the influence of either rotational SRS, the optical Kerr effect (instantaneous nonlinear refractive index), or both of them combined. Interestingly, numerical simulations suggest that by combining the vibrational and rotational response, it is possible for the continuum to form by the Raman effect alone. This is an alternative regime of supercontinuum formation in optical fibers---conventional routes, such as soliton self-compression and fission dynamics~\cite{Husakou2001}, modulational instability \cite{tai_observation_1986}, or self-phase modulation (SPM) and self-steepening, are all dominated by the contribution of the optical Kerr effect. Our numerical results suggest that our approach can produce supercontinuum spectra spanning from the ultraviolet to the mid-infrared (270~nm to beyond 2000~nm). Experimentally we generate a flat supercontinuum spanning 440~nm to 1200~nm and an additional deep ultraviolet continuum from 250~nm to 360~nm. These results are achieved by pumping a nitrogen-filled anti-resonant fiber with 20~ps pump pulses at 532~nm.

\section{Theory and numerical simulations}
We model the propagation of the electric field through the HC-PCF using a unidirectional pulse propagation equation~\cite{Kolesik2004,travers2019high},
\begin{equation}
\begin{split}
    \partial_z\tilde{E}(z,\omega)=i&\left[\beta(z,\omega)-\frac{\omega}{v(z)}\right]\tilde{E}(z,\omega)\\&+\frac{i\omega^2}{2c^2\epsilon_0\beta(z,\omega)}\tilde{P}(z,\omega),
\end{split}
\end{equation}
where $\tilde{E}(z,\omega)=\mathcal{F}[E(z,t)]$ is the Fourier transform of the linearly polarized, mode-averaged electric field $E(z,t)$, $z$ is axial propagation distance through the fiber, $\omega$ is angular frequency, $\beta(z, \omega)$ is the frequency- and pressure-dependent propagation constant of the fundamental fiber mode, which we model using the hollow-capillary model~\cite{marcatili_hollow_1964,travers2019high} with the pressure-dependent dispersion of the gas~\cite{Borzsonyi:08}, $v(z)$ is the group velocity of the reference frame, $c$ is the vacuum light speed, $\epsilon_0$ is the permittivity of free space, and $\tilde{P}(z,\omega)$ is the Fourier transform of the nonlinear polarization
\begin{equation}
    P(z,t) = P^\mathrm{k}(z,t) + P^\mathrm{i}(z,t) + P^\mathrm{v}(z,t) + P^\mathrm{r}(z,t).
\end{equation}
The Kerr polarization is $P^\mathrm{k}(t)=N\gamma^{(3)}E^3$ where $\gamma^{(3)}$ is the third-order hyperpolarizability and $N$ is the number density of molecules. The polarization due to ionization and the plasma response, $P^\mathrm{i}(t)$, is calculated using the method of Geissler~\cite{Geissler1999}, using the tunnelling limit of the Peremolov-Popov-Terent'ev ionization rate~\cite{Perelomov1966} with an ionization potential of 15.58~eV. The vibrational polarization $P^\mathrm{v}$ and the rotational response $P^\mathrm{r}$ will be discussed below. The numerical parameters we use are given in Table~\ref{tab:param}. We neglect the loss of the fiber and the dispersion spikes arising due to the fiber resonances~\cite{Tani:18}.

\begin{table}[b]
\centering
\caption{\label{tab:param}%
Parameters for numerical simulations.}
\begin{tabular}{llrc}
\hline\hline
\textrm{Parameter}&
\textrm{Unit}&
\textrm{Value}&
\textrm{Ref.}\\
$\gamma^{(3)}$ & $10^{-64}$~Cm$^4$V$^{-3}$  & 95.2 & \cite{LEHMEIER198567}\\
$\nu_v$ & THz & 69.85 & \cite{Bendtsen1974} \\
$\Delta \nu_v$ & THz & $\sim\!0.04$ & \cite{koszykowski1987theoretical} \\
$\partial \alpha/\partial Q$ & $10^{-20}$~m$^2$ & 1.75 & \cite{Murphy1969} \\
$\mu$ & $10^{-26}$~kg & 1.16 & \\
$B$ & m$^{-1}$ & 199 & \cite{Bendtsen1974} \\
$D$ & $10^{-4}$~m$^{-1}$ & 5.76 & \cite{Bendtsen1974} \\ 
$q_J$ odd & & 1 & \\
$q_J$ even & & 2 & \\
$\Delta \alpha$ & $10^{-31}$~m$^{3}$ & 6.7 & \cite{Wahlstrand2012} \\
$b_r$ & GHz amagat$^{-1}$ & $\sim\! 3.3$ & \cite{Herring1986} \\
\hline\hline
\end{tabular}
\end{table}

\begin{figure*}
    \centering
    \includegraphics{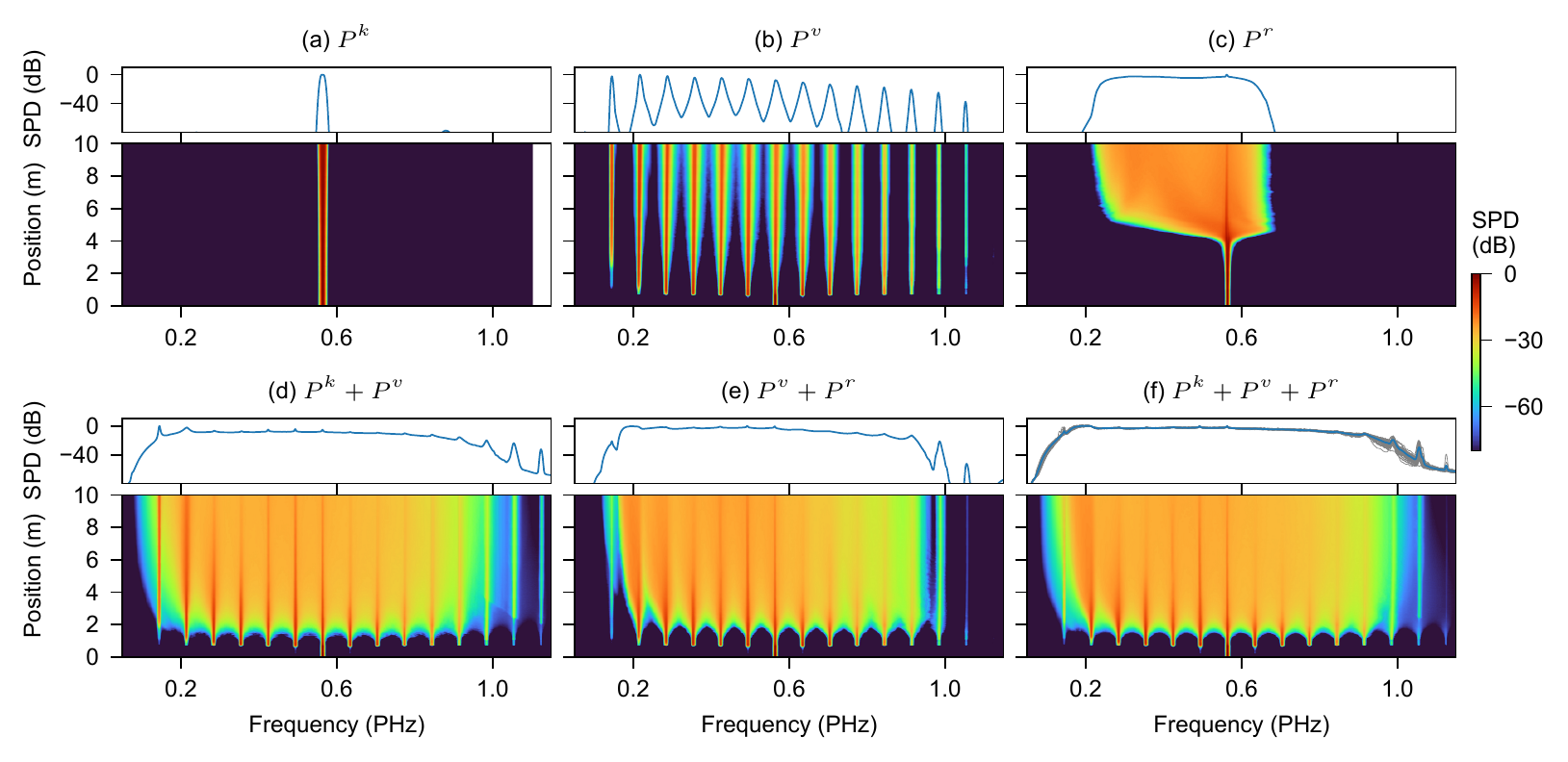}
    \caption{Numerical simulations of the spectral evolution. We model 20~ps duration, 532~nm wavelength, 80~$\upmu$J pump pulses in a 10~m long, 26~$\upmu$m diameter hollow fiber filled with nitrogen with a pressure gradient from 1 to 40~bar. Each subplot includes the indicated contributions to the nonlinear polarization. SPD = relative spectral power density. These data are an ensemble average of 100 simulations to account for shot-to-shot fluctuations. To illustrate the shot-to-shot fluctuations the top panel of (f) shows each individual shot as a thin grey line, with the average in blue.}
    \label{fig:specs}
\end{figure*}

\begin{figure}
    \centering
    \includegraphics{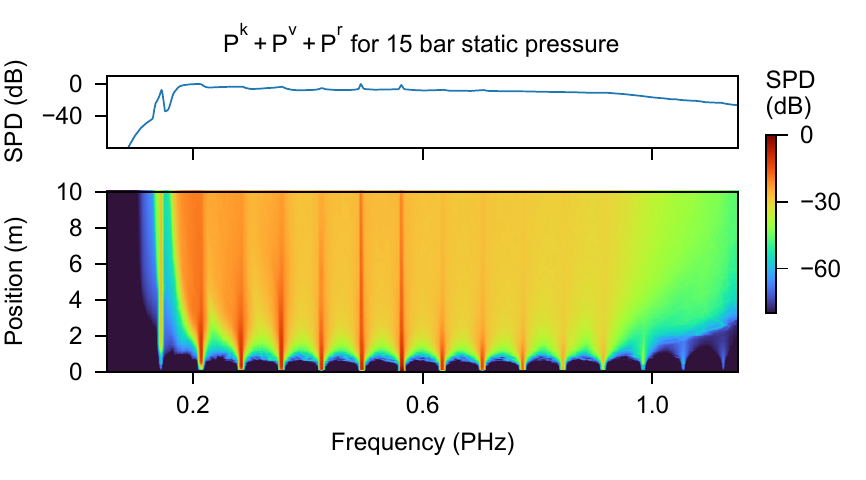}
    \caption{Numerical simulation of the spectral evolution for a constant static pressure of 15~bar in the fiber, instead of a pressure gradient. All other parameters are identical to Fig.~\ref{fig:specs}(f).}
    \label{fig:static}
\end{figure}

Figure~\ref{fig:specs}(a) shows numerical simulation results when including only the Kerr effect (only $P^\mathrm{k}$ in the model) for a 26~$\upmu$m diameter hollow fiber and pulses with 20~ps duration, 80~$\upmu$J energy and a wavelength of 532~nm. We considered a nitrogen pressure gradient from 1 to 40~bar, corresponding to our experimental parameters, which are detailed later. Additional numerical simulations (see Fig.~\ref{fig:static} and discussion around it) show that none of the dynamics we describe here depend on the pressure gradient, and would also occur in a statically filled fiber. Pure Kerr propagation without the Raman contribution results in negligible broadening, because the pump pulse and fiber parameters do not lead to either significant self-phase modulation or modulational instability. The spectral width induced by SPM is approximately given by $\Delta\omega\approx \phi_{\mathrm{max}}/\tau_0$, with $\phi_{\mathrm{max}}=(2/3) n_2\omega P_0L/c A_\mathrm{eff}$, where $n_2=3 \bar{N}\gamma^{(3)}/4\epsilon_0^2c$ is the nonlinear refractive index of the gas, $P_0$ and $\tau_0$ are the peak power and duration of the pulse, $L$ is the fiber length, $A_\mathrm{eff}$ is the effective mode area of the fiber core, $\bar{N}$ is the average number density, and the factor $2/3$ accounts for the pressure gradient. For our parameters this gives $\Delta\nu = \Delta\omega/2\pi \approx 2.6$~THz, and as a consequence the spectrum barely changes.

The generation of a Raman frequency comb from a single pump pulse can be initiated spontaneously from noise~\cite{Raymer1980,Wang2010b} and subsequently amplified through SRS \cite{Couny1118}. The (nonlinear or inelastic) coupling of the light field with the stretching and alignment of homonuclear diatomic molecules is described through first-order perturbations of a non-rigid rotor model~\cite{Wahlstrand2015,Palastro2012,Wahlstrand2013,Chen:07,Nibbering:97}. As the vibrational frequency $\nu_v$ is typically much larger than the rotational frequencies $\nu_r^J$, and large enough that at room temperature only the ground vibrational state is thermally populated, Raman scattering from molecular stretching is captured and modelled by a single vibrational transition. Therefore the time-dependent nonlinear polarization due to the vibrational coupling is described by~\cite{Wahlstrand2015}
\begin{eqnarray}
\label{eqn:vib}
    P^\mathrm{v} &=& E(t)N(4\pi\epsilon_0)^2\kappa_v\int_{0}^{\infty}h(\nu_v, T_2^v, \tau) E(t-\tau)^2d\tau,\\
    \kappa_v &=& -\left(\frac{\partial \alpha }{\partial Q}\right)^2\frac{1}{8\pi \mu \nu_v},
\end{eqnarray}
where the Raman response function is $h(\nu, T_2, t) = \sin (2\pi \nu t)\exp(-t/T_2)$, $T_2=(\pi \Delta\nu)^{-1}$ is the dephasing time with $\Delta\nu$ the full-width at half-maximum (FWHM) linewidth of the Raman transition, $\partial \alpha/\partial Q$ is the isotropic averaged polarizability derivative with respect to the ensemble-averaged molecular stretch $Q$, and $\mu$ is the reduced molecular mass.

\begin{figure}
    \centering
    \includegraphics{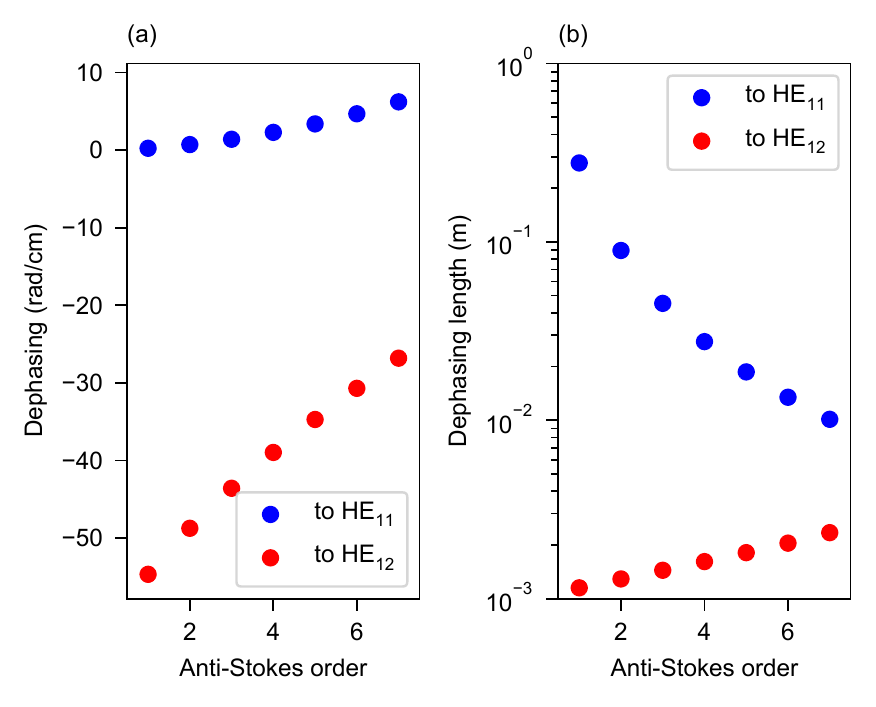}
    \caption{Phase-matching of the anti-Stokes lines in the vibrational frequency comb at a gas pressure of 11~bar (corresponding to the pressure at the point the comb is formed). (a) The phase-missmatch for each anti-Stokes order, calculated as $\Delta\beta = \beta(\omega_{\mathrm{AS}_\mathrm{n}}) - \beta(\omega_{\mathrm{AS}_\mathrm{n-1}}) + \Delta\beta_\mathrm{R}$ where $\omega_{\mathrm{AS}_\mathrm{n}}$ and $\omega_{\mathrm{AS}_\mathrm{n-1}}$ are the frequencies of the current and previous anti-Stokes orders, and $\Delta\beta_\mathrm{R}$ is the propagation constant difference between the pump and first Raman Stokes frequency. (b) The dephasing length of each anti-Stokes order, calculated from (a). For all orders the dephasing length is longer than the distance over which each comb line is generated.}
    \label{fig:phase}
\end{figure}

In Fig.~\ref{fig:specs}(b), we include only the vibrational Raman response $P^\mathrm{v}$. Our parameters correspond to pumping close to the transient gain regime \cite{Raymer1980}, and because the ratio $\Delta\nu_v/\nu_v$ is relatively small (Table~\ref{tab:param}) the Raman response gives rise to a vibrational frequency comb with well-separated peaks at both lower (Stokes) and higher (anti-Stokes) frequencies separated by $\nu_v$. Note that each line results from scattering from the previous line using the same (fundamental) vibrational transition, and hence the Stokes and anti-Stokes lines are equally spaced in frequency. This is similar to previous results in light homonuclear gases such as  hydrogen~\cite{Benabid399, Couny1118, Wang2010, Abdolvand:12, Tani:15, Benoit:15}. As described in some of those previous studies, the smooth dispersion profile of gas-filled fibers allows for phase-matching of the anti-Stokes components of the comb~\cite{Benabid399}. For the parameters used here, vibrational comb formation occurs at around 80~cm, where the gas pressure is 11~bar. The phase mismatch for each Raman comb line is shown in Fig.~\ref{fig:phase}(a). The corresponding dephasing length, $2\pi/\Delta\beta$, is shown in Fig.~\ref{fig:phase}(b). Analysis of the numerical simulations in Fig.~\ref{fig:specs} shows that each comb line forms within just a few centimetres of propagation after the formation of the previous comb line, which is a shorter distance than the dephasing length in all cases apart from the very last comb line. Therefore we can infer that the vibrational comb is indeed phase-matched, and that it is phase-matching which limits the overall extent of the comb, and therefore of the final supercontinuum. Furthermore, conversion to higher order modes has a higher phase mismatch in all cases.

In contrast to the vibrational levels, the spacing of the rotational levels is small enough that a manifold of levels are populated at room temperature and therefore contribute to the nonlinear response. The energy $e_J$ and thermal population $\rho_J$ of each level are given by
\begin{eqnarray}
    e_J&=&hc[BJ(J+1)-DJ^2(J+1)^2],\\
    \rho_J&=&\frac{q_J(2J+1)\exp (-e_J/k_B T)}{\sum_K q_K(2K+1)\exp (-e_K/k_B T)},
\end{eqnarray}
where $J$ is the rotational quantum number, $h$ is Planck's constant, $B$ and $D$ are the rotational and centrifugal constants for the specific molecular species, $k_B$ is Boltzmann's constant, $T$ is the temperature, and $q_J$ is a statistical weighting factor that varies with odd and even $J$ to account for nuclear spin (see Table~\ref{tab:param}). Selection rules limit the rotational transitions to pairs of levels with $\Delta J=2$. In nitrogen we can neglect the centrifugal contribution ($B/D>10^{5}$), and the rotational line spacing is $\nu_r^J = (e_{J+2}-e_J)/h=4cB(J+3/2)$ with $4cB=0.24$~THz. Over 20 rotational levels play a role and the linewidth of each level is pressure-broadened. The Lorentzian gain peaks associated with each rotational state have been found to have a linewidth $\Delta\nu_r = (\pi T_2)^{-1} = \rho b_r$ where $\rho$ is the gas density in Amagat and $b_r$ is an experimentally determined constant~\cite{Herring1986}. For nitrogen at high pressure, the rotational Raman gain spectrum is continuous: at 40~bar, $\Delta\nu_r=0.15$~THz, which is similar to the line spacing $\nu_r^J$~\cite{mikhatlov1959,Jammu1966}.

The rotational Raman polarization is approximated by a sum over all allowed rotational transitions~\cite{Wahlstrand2015},
\begin{eqnarray}
    \nonumber
    P^\mathrm{r} &=& E(t)N(4\pi\epsilon_0)^2\\
    &&\times\sum_{J}\kappa_r^J\int_{0}^{\infty}h(\nu_r^J, T_2^r, \tau)E(t-\tau)^2d\tau,\\
    \kappa_r^J &=& \frac{4\pi\Delta\alpha^2}{15 h}\frac{(J+1)(J+2)}{2J+3}\left(\frac{\rho_{J+2}}{2J+5}-\frac{\rho_J}{2J+1}\right)
\end{eqnarray}
where $\Delta\alpha$ is the molecular polarizability anisotropy.
\begin{figure*}
    \centering
    \includegraphics{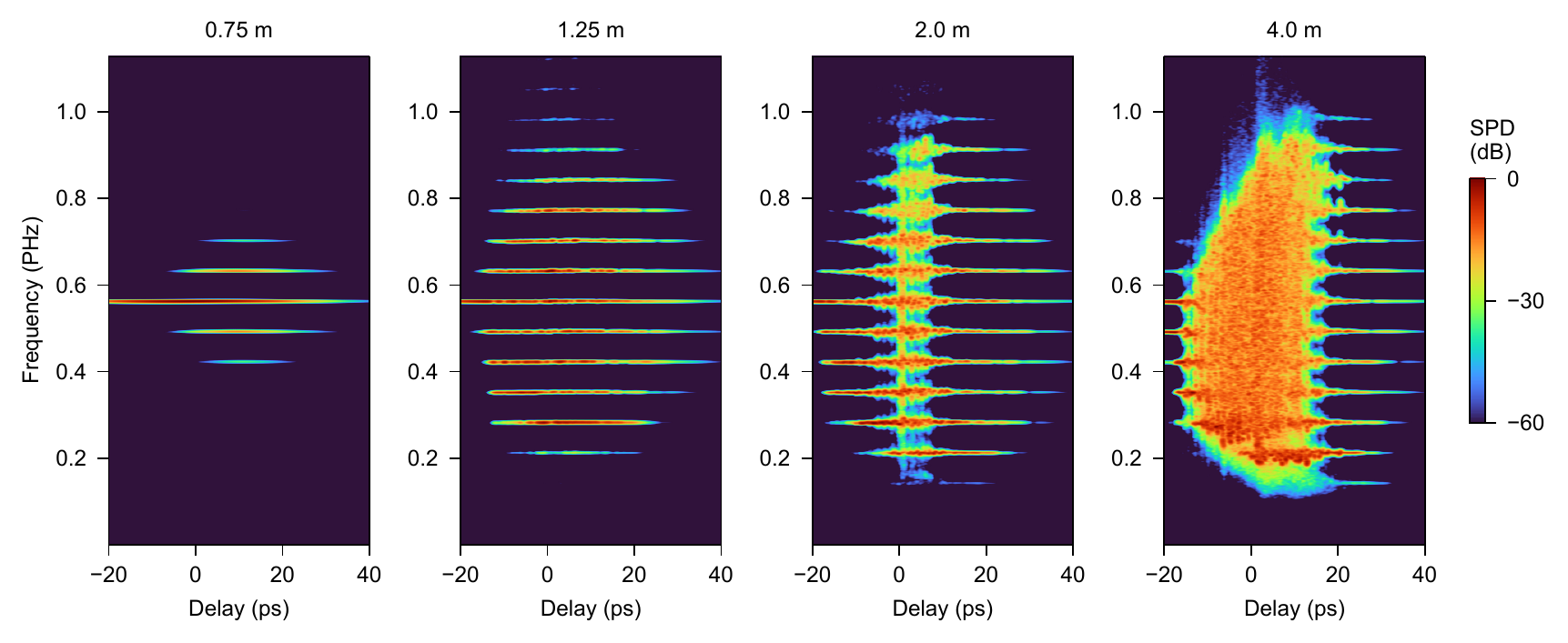}
    \caption{Numerically simulated spectrograms of the evolution for the same parameters as Fig.~\ref{fig:specs} and Fig.~\ref{fig:It} with all contributions included. Each subplot is for a different propagation distance.}
    \label{fig:sgm}
\end{figure*}

\begin{figure}
 \centering
 \includegraphics{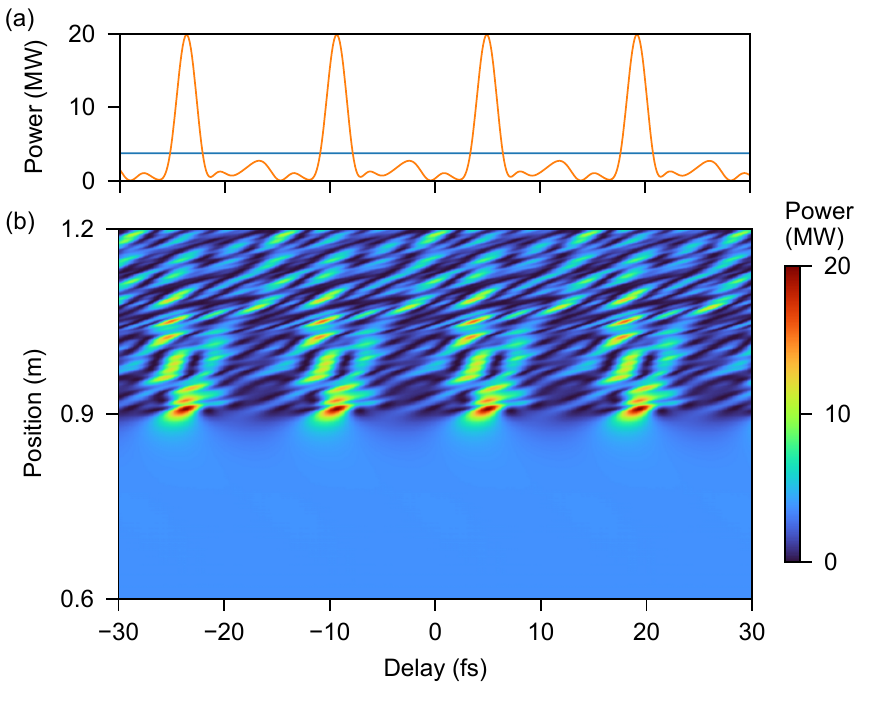}
 \caption{Initial temporal dynamics. We focus only on the central 60~fs of the 20~ps pulse, for the same parameters as Fig.~\ref{fig:specs} with all contributions included (nearly identical results are obtained for pure vibrational evolution for the propagation length shown here).  (a) Slices of the instantaneous power at the input (blue) and after 0.91~m (orange), where the peak power is highest (20~MW) and the pulse duration is 2.1~fs. (b) The instantaneous power evolution from 0.6~m to 1.2~m.}
 \label{fig:It}
\end{figure}

The spectral evolution obtained for pure rotational SRS is shown in Fig.~\ref{fig:specs}(c). It shows different behaviour to the case of pure vibration: the spectrum appears very smooth and continuous, and mostly down-shifted to longer wavelengths; the overall spectral coverage is also much reduced. The frequency down-shift results from the slower rotational SRS response, resulting in an asymmetric, positive, phase-shift as described in detail elsewhere~\cite{Nibbering:97}. This is in contrast to the pure vibrational SRS case, where the pulse experiences an almost pure sinusoidal phase modulation, resulting in a frequency comb.

When combining the vibrational Raman and Kerr effects, the vibrational comb lines broaden into each other (Fig.~\ref{fig:specs}(d)), forming a broadband continuum. Fig.~\ref{fig:sgm} shows spectrograms of the key stages of evolution: vibrational comb formation, followed by spectral broadening of the comb lines into a supercontinuum. As noted previously, the pump pulse does not drive significant SPM. However, in the time domain the vibrational frequency comb corresponds to an ultrashort pulse train. Fig.~\ref{fig:It} shows that after 0.91~m propagation, the vibrational comb consists of a train of $2.1$~fs pulses with a peak power of $20$~MW. The formation of such ultrafast pulse structures, due to the coherent combination of all the Raman comb lines, is consistent with previous studies~\cite{baker_femtosecond_2011,chan_synthesis_2011}. The combination of shorter duration and higher peak power enhances self-phase modulation by a factor of $\sim\! 50000$ compared to the 20~ps pump pulses with a peak power of 3.8~MW, leading to spectral broadening of the comb. Furthermore, the individual comb lines broaden into each other to form a continuum. Previous Raman frequency comb experiments have either worked with much lower peak power and longer pump pulses~\cite{Couny2007b}, producing a pure frequency comb in which SPM never occurs and no continuum is formed, or with much shorter pulses, leading to immediate SPM broadening of the pump pulse itself~\cite{Belli:15}. With our parameters, in contrast, SPM occurs only after the formation of the frequency comb.

While broadening of the vibrational comb due to the Kerr effect leads to a continuum, it is not the only mechanism that can achieve this. Remarkably, combining both vibration and rotation, but not the Kerr effect, as shown in Fig.~\ref{fig:specs}(e), leads to a \textit{pure Raman supercontinuum} spanning multiple octaves achieved via spectral broadening and merging of several vibrational lines. This works well in nitrogen, because the spacing and linewidth of the rotational levels are similar, which creates a near-continuous rotational Raman gain spectrum. However, we expect that this is a universal feature of molecular gases. Our numerical results indicate that in hydrogen (not shown), where the rotational gain spectrum consists of well-separated narrow lines, a smooth supercontinuum can still be formed.

The similar broadening of the vibrational Raman comb into a supercontinuum by either the Kerr or rotational Raman response is not due to any similarity of those processes, but because both effects phase-modulate the ultrafast pulse structures underlying the vibrational comb (Fig.~\ref{fig:It}) to varying degrees based on their peak power and temporal location. It is this non-constant phase modulation which leads to broadening and merging of the vibrational comb lines, and this is independent of the specific physical mechanisms involved.

Figure~\ref{fig:specs}(f) shows the full supercontinuum simulations, including all contributions to the nonlinear polarization. Importantly it is the vibrational comb that establishes the full continuum extent, covering 3 octaves, and then the Kerr and rotational Raman responses that broaden the vibrational lines to form a true continuum.

Ionization plays a negligible role in the dynamics---the low peak intensity (a maximum of $3\times 10^{13}$~W/cm$^2$ is reached) leads to a negligible ionisation fraction of 0.002\%. This is in contrast to supercontinuum generation in gas-filled hollow fibers pumped with shorter pulses based on modulational instability~\cite{Tani2013}, where ionization contributes significantly.

The supercontinuum spectra shown in Fig.~\ref{fig:specs} are an average of 100 simulations to account for shot-to-shot fluctuations. Like other noise seeded supercontinuum mechanisms, such as modulational instability, each laser shot produces a slightly different spectrum. To illustrate these fluctuations the top panel of Fig.~\ref{fig:specs}(f) shows each individual shot as a thin grey line, with the average in blue. These fluctuations are not detrimental to most applications, which acquire time-averaged spectra.

In the numerical simulations described so far we used a gas pressure gradient from 1 to 40~bar, corresponding to our experiments. To check whether the pressure gradient is critical to the dynamics shown here, Fig.~\ref{fig:static} shows a simulation for a static gas pressure of 15~bar, with all other parameters are identical to Fig.~\ref{fig:specs}(f). The overall dynamics are virtually identical, confirming that the supercontinuum mechanism we have identified does not depend on a specific gas pressure or gradient. The difference in spectral extent at high frequencies is due to variations in phase-matching.

\section{Experiment}
We verified these dynamics experimentally using a single-ring nodeless hollow-core anti-resonant fiber consisting of 6 untouching thin tubes with diameter of 12~$\upmu$m and membrane thickness of 210~nm forming a negative-curvature core shape with inscribed diameter of 26~$\upmu$m. The fiber exhibits a minimum attenuation of 80~dB/km at the pump laser wavelength of 532~nm and its first-order transmission band spans from 440~nm to $>1200$~nm with single-mode operation~\cite{Gao2017}. The ultraviolet transmission band was not measured, but results from finite element modelling indicate that it spans 250 nm to 360 nm. We used a 10~m length of fiber, and for experimental convenience we left  the input end open to atmosphere, while keeping the output end sealed inside a gas cell filled with 40~bar of nitrogen. Our 532~nm pump laser emitted 20~ps duration pulses at 1~kHz, and was coupled to the fiber through a telescope and plano-convex lens. The output spectrum was collected by an integrating sphere and two optical spectrum analyzers. 
\begin{figure}
    \centering
    \includegraphics[width=8.6cm]{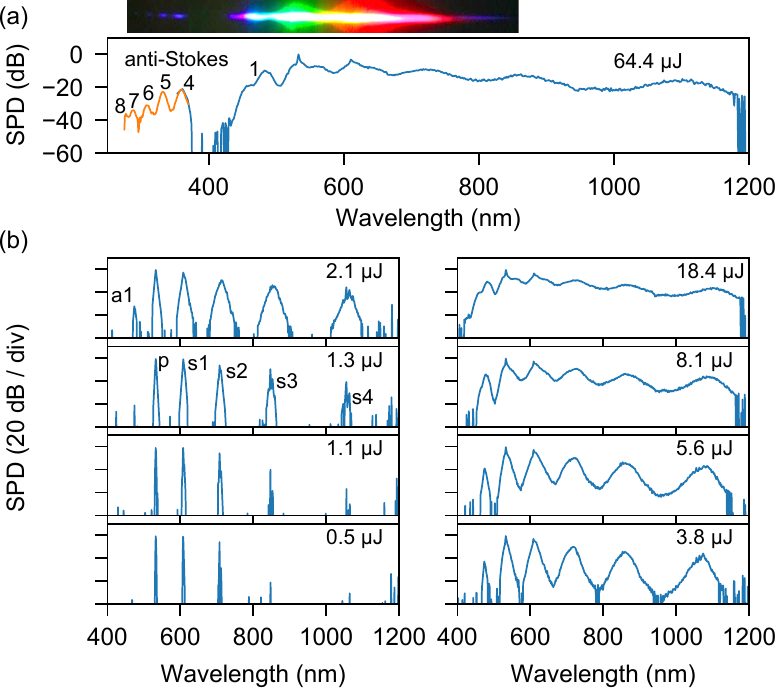}
    \caption{Experimental measurements. (a) The measured spectrum at the highest pulse energy of 64.4~$\upmu$J. The orange part of the spectrum was measured with a dedicated DUV spectrometer. The inset shows a photograph of the dispersed spectrum. The vibrational anti-Stokes lines are labelled at 426~nm (1), 356~nm (4), 331~nm (5), 305~nm (6), 287~nm (7) and 269~nm (8) (b) The spectral evolution as the input pulse energy is gradually increased from 0.5~$\upmu$J to 18.4~$\upmu$J as indicated. The pump line is labelled (p) as are the discrete vibrational Stokes lines at 607~nm (s1), 707~nm (s2), 847~nm (s3), and 1056~nm (s4).}
    \label{fig:escale}
\end{figure}

Figure~\ref{fig:escale}(a) shows the measured spectrum at the highest pulse energy of 65~$\upmu$J coupled into the fiber. A wide supercontinuum spectrum is generated, in agreement with our numerical simulations, spanning more than 775~nm from 440~nm to 1200~nm (the long-wavelength limit of our measurements). Furthermore, we observe deep ultraviolet (DUV) generation from 260~nm to 360~nm. The presence of the DUV component is further confirmed by the photograph of the dispersed spectrum shown as an inset. The spectral gap from 375~nm to 440~nm is due to the high resonance loss in the hollow-core fiber~\cite{Gao2017}, resulting in the absence of the 2$^\mathrm{nd}$ and 3$^\mathrm{rd}$ vibrational anti-Stokes (AS) lines (426~nm and 387~nm). This could be minimized through further optimization of the uniformity of the anti-resonant tubes of the fiber.

Figure~\ref{fig:escale}(b) shows the output spectral evolution as the input pulse energy is gradually increased from 0.5~$\upmu$J to 18.4~$\upmu$J. At low pulse energies, discrete vibrational Stokes lines at 607~nm (s1), 707~nm (s2), 847~nm (s3), and 1056~nm (s4) are observed. A further increase of the input pulse energy shows spectral broadening of the vibrational lines. Increasing the pulse energy also results in the generation of an anti-Stokes vibrational line at 473~nm (a1). That these lines require higher energies can be attributed to the fiber loss.  All of the vibrational lines generated experience spectral broadening and merge to form the supercontinuum at around 8~$\upmu$J of pump energy. The DUV part of the continuum shown in Figure~\ref{fig:escale}(a) corresponds to vibrational anti-Stokes lines at 356~nm (a4), 331~nm (a5), 305~nm (a6), 287~nm (a7), 269~nm (a8) and their subsequent spectral broadening. These dynamics are in excellent agreement with our numerical modelling.

\section{Discussion}
Experiments in molecular gases in HC-PCF to date have mostly concentrated on pumping with pulses of much longer duration than the Raman oscillation period. Apart from comb generation, this system has also been used for extremely efficient Raman conversion and generation~\cite{Benabid2004,Benabid2004a}, continuous-wave pumped gas-Raman lasers~\cite{Couny2007b}, solitary pulse generation~\cite{Abdolvand2009}, self-similar evolution~\cite{Nazarkin2010} and supercontinuum up-conversion~\cite{Bauerschmidt:14}. With much shorter pump pulses, Raman effects combine with a strong Kerr contribution, allowing for enhanced supercontinuum generation~\cite{Debord:19}. In the impulsive regime, with a pump pulse duration close to the Raman oscillation period, Raman dynamics combined with Kerr-driven soliton self-compression have enabled supercontinuum generation from the vacuum ultraviolet to the near infrared~\cite{Belli:15}, ultrafast coherent pulse shaping of a dispersive wave in the ultraviolet~\cite{Belli2018}, and asymmetrical spectral broadening for pulse post-compression~\cite{Beetareabb5375}.

The use of intermediate pulse durations of a few picoseconds leads to completely different nonlinear dynamics characterized by efficient noise-seeded generation of vibrational lines and a frequency comb followed by spectral broadening and supercontinuum generation. Recently a single frequency-shifted vibrational line was generated by pumping an atmospheric air-filled HC-PCF with 6~ps pulses~\cite{Mousavi:18}, and this line broadened, along with the pump spectrum, into a high-power sub-octave continuum. Here we have shown that even a broad frequency comb, consisting of a multitude of vibrational lines (we experimentally demonstrate 10 vibrational lines, numerically more than 12) can be broadened into a flat continuum. This dramatically increases the width of the continuum to cover multiple octaves. For the fiber parameters, gas species and pressure that we use, the pump pulses cannot drive Kerr-based broadening mechanisms, such as self-phase modulation, soliton dynamics or modulational instability, by themselves. Instead, it is only through the sequential effects of vibrational comb formation and consequent breakup of the long pump pulse into few-femtosecond pulses, which then proceed to drive both rotational Raman scattering and self-phase modulation, that the continuum is created.

We expect our observations to be a general result for other molecular gases too, albeit with different characteristics. This scheme could be expanded for VUV and mid-infrared supercontinuum generation by altering the pump wavelength and using a fiber with suitably tuned resonances. Furthermore, this technique may be scalable to high average power by increasing the repetition rate. Research in that direction must carefully consider possible heating of the gas due to excitation through Raman scattering.

\section{Acknowledgements}
This work is supported by the National Natural Science Foundation of China (No. 61827820, 62105122), Guangdong Basic and Applied Basic Research Foundation (No. 2021B1515020030, 2021A1515011646), and the Fundamental Research Funds for the Central Universities (21620414). JCT, FB and CB are funded by the European Research Council (ERC) under the European Union's Horizon 2020 research and innovation program: Starting Grant agreement HISOL, No. 679649. This work used EPCC's Cirrus HPC Service (https://www.epcc.ed.ac.uk/cirrus).

\providecommand{\noopsort}[1]{}\providecommand{\singleletter}[1]{#1}%

\end{document}